\documentclass[10pt]{article}
\usepackage[dvips]{graphicx}
\begin{document}
\begin{center}
\textbf{{\large Parametrization of Born-Infeld Type Phantom Dark
Energy Model}}
 \vskip 0.35 in
\begin{minipage}{4.5 in}
\begin{center}
{\small Z. G. HUANG$^{1,~\dag}$ \vskip 0.06 in \textit{ $^1$School
of Science,
\\~Huaihai~Institute~of~Technology,~222005,~Lianyungang,~China
\\
$^\dag$zghuang@hhit.edu.cn}} \vskip 0.25 in {\small H. Q.
LU$^{2,~\ddag}$ and  W. FANG$^{2}$ \vskip 0.06 in \textit{
$^2$Department~of~Physics,~Shanghai~University,~Shanghai,~China
\\
$^\ddag$alberthq$\_$lu@staff.shu.edu.cn}}
\end{center}
\vskip 0.2 in

{\small Applying the parametrization of dark energy density, we can
construct directly independent-model potentials. In Born-Infeld type
phantom dark energy model, we consider four special parametrization
equation of state parameter. The evolutive behavior of dark energy
density with respect to red-shift $z$, potentials with respect to
$\phi$ and $z$ are shown mathematically. Moreover, we investigate
the effect of parameter $\eta$ upon the evolution of the constructed
potential with respect to $z$. These results show that the evolutive
behavior of constructed Born-Infeld type dark energy model is quite
different from those of the other models.\vskip 0.2 in
\textit{Keywords:} Dark energy; Born-Infeld Field; Parametrization;
Potential.
}
\end{minipage}
\end{center}
\vskip 0.2 in
\begin{flushleft}\textbf{1. Introduction}\end{flushleft}

Since the first observational data of SNe Ia was issued in 1998[1],
the astrophysics and cosmology have undergone a deep change and were
facing a grant challenge. Up to now, series of astrophysical
observational data such as 5 year WMAP$[2], SDSS$[3] together with
SNe Ia data show us such a fact clearer and clearer: the Universe is
spatially flat to high precision, $\Omega_{total}=0.99-0.03$[4], and
consists of about two thirds unknown energy density, one third dust
matter including cold dark matters plus baryons, and negligible
radiation, and that our universe is undergoing an accelerated
expansion. This unknown energy density which is called "dark energy"
with negative pressure pervades the whole university and is
unclumped. We know little on the nature of dark energy thought many
models of dark energy such as cosmological constant model[5-10],
quintessence[11-17], tachyon[18-21], holographic dark energy[22],
K-essence model[23-25] and phantom model[26], have been proposed and
investigated. Thus phenomenological investigation are good choices
and have been widely conducted by parametrizing the dark energy
equation of state(EOS) $\omega$. To study the evolution of the
universe, different kinds of potentials can be put into different
model with scalar field and then the EOS $\omega$ would be studied.
Of course, potentials can also be reconstructed from a
parametrization of the EOS $\omega$ fitting current observational
data. The latter looks more persuasive to find out a best-fit dark
energy model, because it is independent on model.
\par Born-Infeld(shortened B-I) field was firstly introduced to explain the
singularity in classical electromagnetic dynamics by Born and Infeld
in 1934[27]. So far, many authors have been studying the nonlinear
B-I type string theory and cosmology. Their work show that the
lagrangian density of this B-I type scalar field posses some
interesting characteristics[28]. Recently, we investigated the
universe of B-I type scalar field with potential and find that this
model can undergo a phase of accelerating expansion corresponding
EOS $-1<\omega<-\frac{1}{3}$[29]. This model admits a late time
attractor solution that leads to EOS $\omega=-1$. The lagrangian of
B-I type scalar field with negative kinetic energy also is
considered by us. An interesting result is that weak energy
condition and strong energy condition are violated for phantom B-I
type scalar field. In this model, the EOS $\omega$ is always smaller
than -1, which meets the current observation data well.
\par Guo et al.[30] have constructed a theoretical method of constructing
the quintessence potential $V(\phi)$ directly from the dark energy
equation of state function $\omega$. In Ref.[31], we apply the
method of parametrization of dark energy density function to the
dilaton coupled quintessence(DCQ) model. According to the
comparision between the constructed DCQ potential and quintessence
potential($\alpha=0$), we find that the shapes of the constructed
DCQ potential quite different from the one of quintessence
potential. An interesting result is DCQ potentials possess two
different evolutive mode $"O"$ and $"E"$.
\par In this paper we will investigate the
parametrization of B-I type phantom dark energy model. Using the
model-independent method, we construct potential which is best-fit
observations. Many authors have presented various parametrization of
EOS $\omega_\phi$ of dark energy and investigated their interesting
features. In this paper we choose four cases of them to reconstruct
the potential of B-I type dark energy directly from the EOS
$\omega_\phi$. The four typical parametric of EOS :
$\omega_\phi=\omega_0$; Case II: $\omega_\phi=\omega_0+\omega_1z$;
Case III: $\omega_\phi=\omega_0+\omega_1\frac{z}{1+z}$; Case IV:
$\omega_\phi=\omega_0+\omega_1ln(1+z)$, have been proved that they
fit the observations well in different areas of red-shift $z$. The
numerical results of the evolution of scalar potential with respect
to scalar field $\phi$ and red shift $z$ will be shown in the plots
correspondingly. The evolutions of the dark energy density
$\rho_\phi$  with respect to $z$ are also shown mathematically. This
paper is organized as follows: Section I is introduction. In Section
II, we introduce the equations of B-I type field firstly, after
reconstructing the scalar potential of B-I type dark energy model by
parametrizing the EOS $\omega_\phi$, we consider four special cases
and get the corresponding mathematically results. Section III is
summary.

\vskip 0.2 in
\begin{flushleft}\textbf{2. Basic Equations and Numerical Results}\end{flushleft}
\par
The lagrangian density for a B-I type phantom scalar field is
\begin{equation}\displaystyle L_S=\frac{1}{\eta}\left[1-\sqrt{1+\eta g^{\mu\nu}\phi_{,~\mu}\phi_{,~\nu}}~\right]\end{equation}
Eq.(1) is equivalent to the tochyon lagrangian
$[-V(\phi)\sqrt{1+g^{\mu\nu}\phi_{,~\mu}\phi_{,~\nu}}+\Lambda]$ if
$\displaystyle V(\phi)=\frac{1}{\eta}$ and cosmological constant
$\displaystyle \Lambda=\frac{1}{\eta}$ ($\displaystyle
\frac{1}{\eta}$ is two times as "critical" kinetic energy of $\phi$
field). Now we consider the Lagrangian with a potential $V(\phi)$ in
spatially homogeneous scalar field, Eq.(1) becomes
\begin{equation}\end{equation}$$L_S=\frac{1}{\eta}\left[1-\sqrt{1+\eta \dot{\phi}^2}~\right]-V(\phi)$$
In the spatially flat Robertson-Walker metric
$ds^2=dt^2-a^2(t)(dx^2+d^2y+d^2z)$, Einstein equation
$G_{\mu\nu}=KT_{\mu\nu}$, can be written as
\begin{equation}H^2=\frac{1}{3}(\rho_\phi+\rho_m)\end{equation}
\begin{equation}\frac{\ddot{a}}{a}=-\frac{1}{6}(\rho_{\phi}+3p_{\phi}+\rho_m)\end{equation}
\begin{equation}\ddot{\phi}+3H\dot{\phi}(1+\eta{\dot{\phi}}^2)-V'(\phi)(1+\eta{\ddot{\phi}}^2)^{\frac{3}{2}}=0\end{equation}
\begin{equation}\dot{\rho}_m+3H(\rho_m+p_m)=0\end{equation}
\begin{equation}\dot{\rho}_\phi+3H(\rho_\phi+p_\phi)=0\end{equation}
where $\rho_m$, $\rho_\phi$ and $p_\phi$ are the matter energy
density, the effective energy density and effective pressure of the
B-I type scalar field respectively, the prime sign denotes the
derivative to $\phi$ and we work in units $8\pi G=1$. The
Energy-moment tensor is
\begin{equation}\displaystyle T^\mu_\nu=-\frac{g^{\mu\rho}\phi_{,~\nu}\phi_{,~\rho}}{\sqrt{1+\eta g^{\mu\nu}\phi_{,~\mu}\phi_{,~\nu}}}-\delta^\mu_\nu L_S\end{equation}
From Eq.(8), we have
\begin{equation}\rho_\phi=\frac{1}{\eta\sqrt{1+\eta \dot{\phi}^2}}-\frac{1}{\eta}+V(\phi)\end{equation}
\begin{equation}p_\phi=\frac{1}{\eta}-\frac{\sqrt{1+\eta \dot{\phi}^2}}{\eta}-V(\phi)\end{equation}
From Eqs.(9)(10), we get
\begin{equation}\rho_\phi+p_\phi=\frac{-\dot{\phi}^2}{\sqrt{1+\eta \dot{\phi}^2}}\end{equation}
According to the Eq.(11), we can obtain such a result: when
$\eta\rightarrow0$, B-I type dark energy model reduces to the
$\Lambda$CDM model. In order to construct $V(\phi)$ directly
independent-model, we need to get the analytic expression of
$V(\phi)$ with $\rho_\phi$ and $p_\phi$. We can obtain the analytic
solutions of Eqs.(9) and (10) set,
\begin{equation}V(\phi)=-p_\phi+\frac{1}{\eta}-\frac{1}{\eta}\sqrt{1+\frac{1}{2}\eta(\rho_\phi+p_\phi)[\eta(\rho_\phi+p_\phi)-\sqrt{4+(\eta\rho_\phi)^2+\eta p_\phi)^2+2(\eta)^2\rho_\phi p_\phi}~]}\end{equation}
\begin{equation}(\dot{\phi})^2=\frac{1}{2}(\rho_\phi+p_\phi)[\eta(\rho_\phi+p_\phi)-\sqrt{4+(\eta\rho_\phi)^2+\eta p_\phi)^2+2(\eta)^2\rho_\phi p_\phi}~]\end{equation}
Obviously, when $\eta\rightarrow0$ we get
\begin{equation}\lim_{\eta \to 0}V(\phi)=\frac{\rho_\phi-p_\phi}{2}\end{equation}
\begin{equation}\lim_{\eta \to 0}(\dot{\phi})^2=-(\rho_\phi+p_\phi)\end{equation}
which corresponds to linear phantom scalar field case. In this
model, we consider dark energy and matter including baryon matter
and cold dark matter, and neglect radiation. From Eqs.(3) and (4),
we deduce
\begin{equation}\rho_\phi=3H^2-\rho_{m0}(1+z)^{3}\end{equation}
\begin{equation}p_\phi=-2\dot{H}-3H^2\end{equation}
We define the dimensionless dark energy function $\zeta(z)$ as
follows
\begin{equation}\zeta(z)=\frac{\rho_\phi}{\rho_{\phi_0}}\end{equation}
where $\rho_{\phi_0}$ is dark energy density at red-shift
$z=0$(present). Using Eqs.(2)(18), we get
\begin{equation}\frac{dH}{dz}=\frac{\rho_{\phi_0}}{6H}\frac{d\zeta}{dz}+\frac{\rho_{m_0}(1+z)^2}{2H}\end{equation}
\begin{equation}\dot{H}=-H(z+1)\frac{dH}{dz}\end{equation}
\begin{equation}\dot{\phi}=-H(z+1)\frac{d\phi}{dz}\end{equation}
Substituting Eqs.(16),(17) and (19-21) into Eqs.(12) and (13), we
have
\begin{equation}V(z)=2\dot{H}+3H^2+\frac{1}{\eta}-\frac{1}{\eta}\sqrt{1+\frac{\eta}{2}(-2\dot{H}-\rho_m)\{-\eta[2\dot{H}+\rho_m]-\sqrt{f(z)}\}}\end{equation}
\begin{equation}(\frac{d\phi}{dz})^2=-\frac{1}{2}(2\dot{H}+\rho_m)[-\eta(2\dot{H}+\rho_m)-\sqrt{f(z)}]\end{equation}
where
\begin{equation}f(z)=4+\eta^2(3H^2-\rho_m)^2+\eta^2(2\dot{H}+3H^2)^2-2\eta^2(3H^2-\rho_m)\times(2\dot{H}+3H^2)\end{equation}
\begin{equation}H=H_0E(z)=\sqrt{\frac{\rho_0}{3}}\{[\Omega_{m0}(1+z)^3+(1-\Omega_{m0})(1+z)^{3(1+\omega_\phi)}]\}^{\frac{1}{2}}\end{equation}
\begin{equation}\dot{H}=-H(1+z)\frac{dH}{dz}=\rho_0[-\frac{1-\Omega_{m0}}{6}(1+z)\frac{d\zeta}{dz}-\frac{\Omega_{m0}}{2}(1+z)^3]\end{equation}
\begin{equation}\rho_m=\rho_{m0}(1+z)^3\end{equation}
$\Omega_{m_0}\equiv\frac{\rho_{m0}}{\rho_{0}}$ is the matter energy
density with $\rho_0=\rho_{\phi_0}+\rho_{m_0}$ being present total
energy density, and $E(z)$ is the cosmic expansion rate relative to
its present value. Clearly, when $\eta\rightarrow0$, Eq.(22) and
(23) can be expressed
\begin{equation}\lim_{\eta \to 0}V(\phi)=\dot{H}+3H^2-\frac{1}{2}\rho_{m}\end{equation}
\begin{equation}\lim_{\eta \to 0}(\dot{\phi})^2=-2\dot{H}-\rho_{m}\end{equation}
which expresses the linear case.  Next we investigate the evolutive
properties of the constructed nonlinear B-I type scalar potential
numerically. In this model , we take the simple dimension dark
energy function $\zeta(z)=(1+z)^{3(1+\omega_\phi)}$ where
$\omega_\phi<-1$ corresponding to phantom dark energy model.
\par Now let us consider four cases[32-37],
which fit the observations well.
\par \textbf{Case I}: $\omega_\phi=\omega_0$[31]
\par \textbf{Case II}: $\omega_\phi=\omega_0+\omega_1z$[32]
\par \textbf{Case III}: $\omega_\phi=\omega_0+\omega_1\frac{z}{1+z}$[33-35]
\par \textbf{Case IV}: $\omega_\phi=\omega_0+\omega_1ln(1+z)$[36]
\vskip 0.3 in
\begin{minipage}{0.45\textwidth}
\includegraphics[scale=0.7]{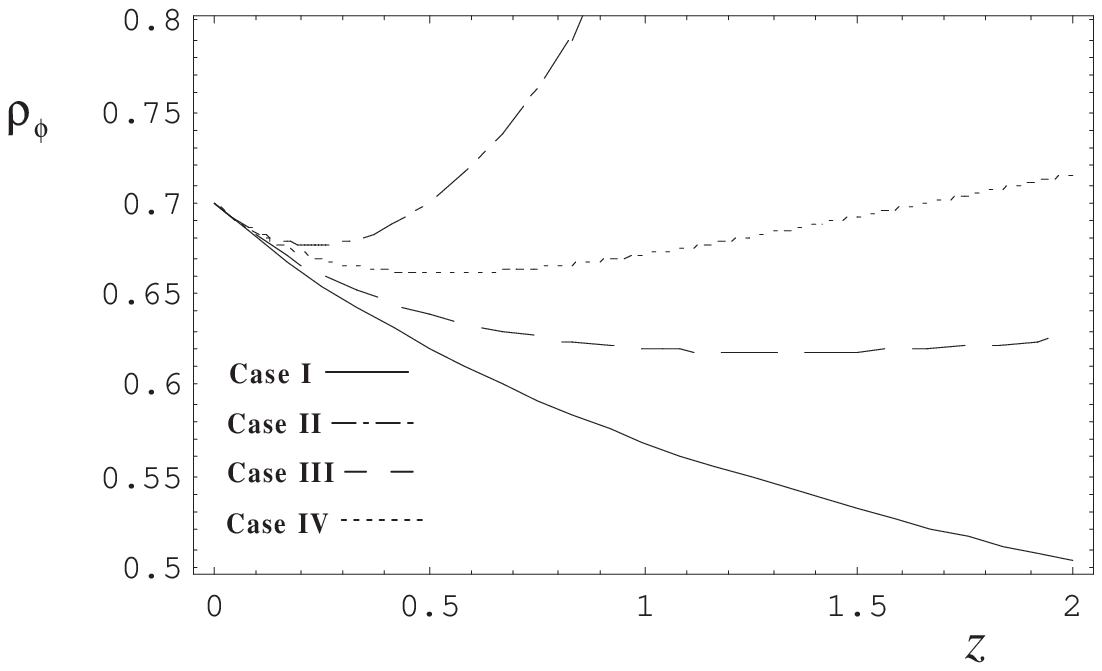}
{\small Fig.1 The evolution of B-I type phantom dark energy density
$\rho_\phi$ with respect to $z$ in the four cases: Case I(real
line), Case II(dot-dashed line),Case III(dot line), Case IV(dashed
line) in B-I type phantom dark energy model. We set
$\Omega_{m_0}=0.3$, $\eta=0.01$, $\rho_0=1$, $\omega_0=-1.1$,
$\omega_1=0.16$.}
\end{minipage}
\hspace{0.08\textwidth}
\begin{minipage}{0.45\textwidth}
\includegraphics[scale=0.7]{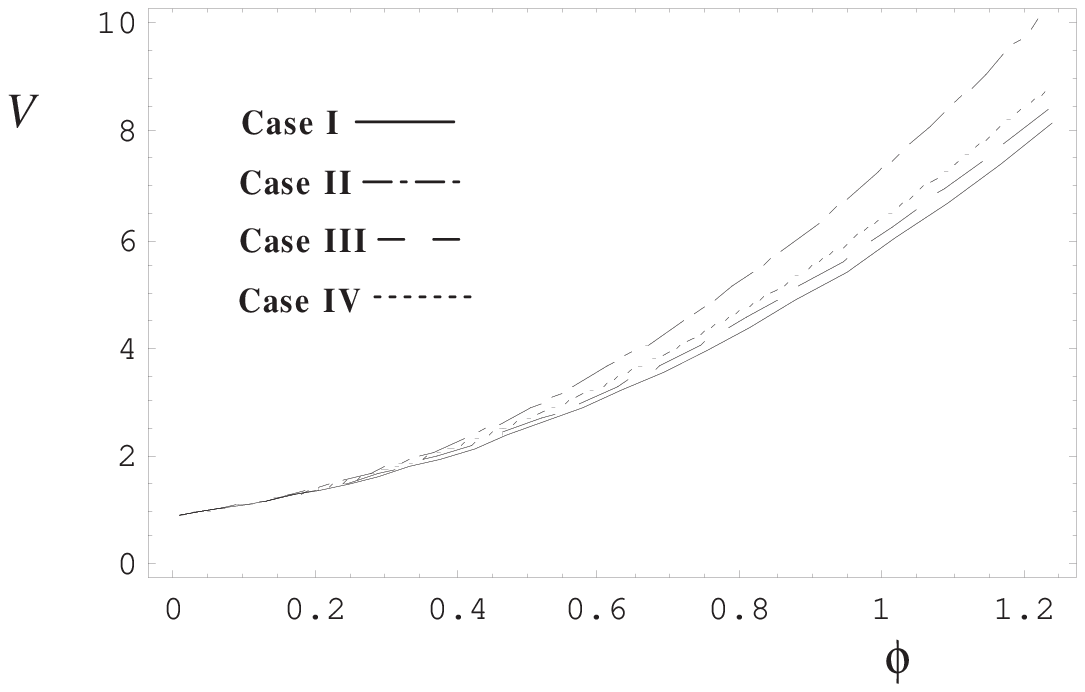}
{\small Fig.2 Constructed B-I type scalar potentials $V$-$\phi$ for
Case I(real line), Case II(dot-dashed line),Case III(dot line), Case
IV(dashed line). We set the same parameter values as those of
Fig.1.}
\end{minipage}
\begin{minipage}{0.45\textwidth}
\includegraphics[scale=0.7]{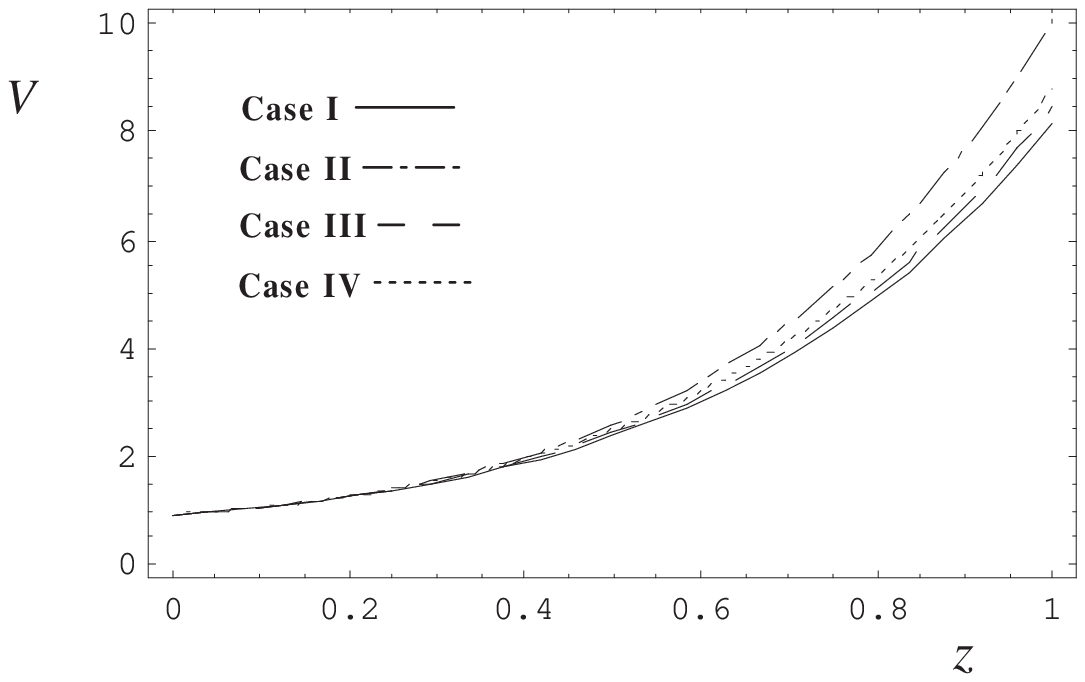}
{\small Fig.3 Constructed B-I type phantom scalar potentials $V$-z
for Case I(real line), Case II(dot-dashed line),Case III(dot line),
Case IV(dashed line). We set the same parameter values as those of
Fig.1.}
\end{minipage}
\hspace{0.08\textwidth}
\begin{minipage}{0.5\textwidth}
\includegraphics[scale=0.7]{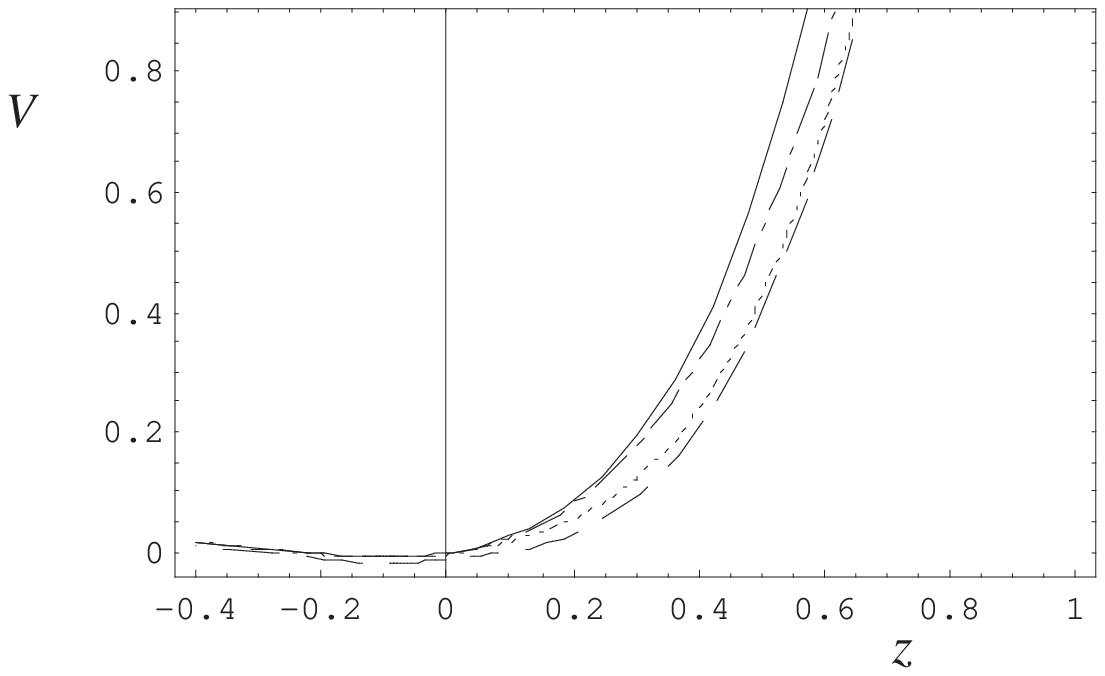}
{\small Fig.4 Constructed B-I type scalar potentials for different
$\eta$: $\eta=0.0001$(real line), $\eta=5$(dot line) and phantom
case $\eta=30$(dot-dash line), $\eta=8000$(dash line). We set
$\Omega_{m0}=0.3$, $\rho_0=0.01$, $\omega_0=-1.1$ and $\omega_1=0$.}
\end{minipage}

\vskip 0.3 in \par We can see from Fig.1 that the evolutional
behaviors of the dark energy density $\rho_\phi$ with respect to $z$
in the four cases tend to be one in the low red-shift
area($0<z<0.15$), on the contrary, very different in the higher
red-shift area($z>0.15$). The constructed potentials $V(z)$ with
respect to $\phi$ by four different parametrizations are
mathematically shown in Fig.2. The evolution of constructed
potentials in Case I, case II and case IV basically keep in step in
the low red-shift areas as not they are in high red-shift. The
evolutions of B-I type scalar potentials with respect to red-shift
$z$ are also plotted in Fig.3. We can see from this figure that in
the case $\omega<-1$ the phantom potentials increases as the
universe expands. This evolutive behavior is quite different from
quintessence potentials whose curve goes downwards. In quintessence
model the system tends to be stable when the potential arrives at
its minimum value, on the contrary, in phantom model the system
tends to be stable when the potential arrives at its maximum value.
So, we see a upward curve before the B-I phantom potential arrives
at its maximum value which corresponds to attractor solution. The
evolutional behavior of constructed potentials for different
$\eta$(where we set $\eta=0.0001,~5,~30$ and $8000$) are shown
mathematically in Fig.4. Obviously, the evolutive trend of
constructed potentials with respect to red-shift $z$ in B-I type
dark energy model are quite similar in low red-shift
area($-0.4<z<0.1$) and become different in higher red-shift
area($z>0.1$).

\vskip 0.3 in
\begin{flushleft}\textbf{{3. Summary}}\end{flushleft}
Using the method of parametrization of dark energy density function
$\zeta(z)$, we construct directly the independent-model phantom
potentials of B-I type dark energy model. In this paper, we consider
the simple parametrization $(1+z)^{3(1+\omega_\phi)}$ . Based on
this, we investigate four special cases which have been widely
studied. Comparising the evolutive orbit of constructed B-I
type(nonlinear) potentials with the linear scalar potentials, we
find that they are basically possess similar characters: the same
behavior in low red-shift areas and quite different in high
red-shift. The evolutive curve of B-I type phantom potentials go
upwards before they arrive at their maximum values, because phantom
field possesses abnormal characters: the kinetic energy of phantom
field is negative. The investigation of effect of different $\eta$
values upon the evolution of B-I type phantom potential with respect
to $z$ shows that the effect is very small in low red-shift area but
becomes larger with the increasing of red-shift values. In the case
of $\eta\rightarrow0$, the nonlinear B-I type phantom dark energy
model reduces to be ordinary phantom model. Recent astrophysical
data[38] show us the(constant) effective equation of state (EOS)
parameter $\omega$ of dark energy should lie in the internal $-1.48
<\omega<-0.72$. For a phantom model the internal becomes $-1.48
<\omega<-1$, therefore our model fits the observational data well.

\begin{flushleft}\textbf{Acknowledgements}\end{flushleft}
This work is partially Supported by the Shanghai Research Foundation
under Grant No 07dZ22020, the Natural Science Foundation of Jiangsu
Province under Grant No 07KJD140011, and the Natural Science
Foundation of HHIT under Grant No Z2007022.

\begin{flushleft}{\noindent\bf References}
 \small{

\item {1.}{ A. G. Riess, \textit{Astron. J}\textbf{116}, 1009(1998).}
\item {2.}{ G. Hinshaw et al., \textit{Astrophys. J. Suppl.}\textbf{180} 225(2009);
\\\hspace{0.15 in}M. R. Nolta et al., \textit{Astrophys. J. Suppl.}\textbf{180} 296(2009).}
\item {3.}{ M. Tegmark et al., \textit{Phys. Rev. D}\textbf{69}, 103510(2004);}
\item {4.}{ P. de Bernardis et al., astro-ph/0105296.}
\item {5.}{ W.J. Percival et al., \textit{Mon. Not. Roy. Astron. Soc}\textbf{337}, 1068(2002).}
\item {6.}{ S. Weinberg \textit{Rev. Mod. Phys.}\textbf{61}, 1(1989).}
\item {7.}{ V. Sahni and A. Starobinsky, \textit{Int. J. Mod. Phys. D}\textbf{9}, 373(2000).}
\item {8.}{ S. M. Carroll, \textit{Living Rev. Rel.}\textbf{4}, 1(2001).}
\item {9.}{ P. J. E. Peebles and B. Ratra, \textit{Rev. Mod. Phys.}\textbf{75}, 559(2003).}
\item {10.}{ T. Padmanabhan, \textit{Phys. Rept.}\textbf{380}, 235(2003).}
\item {11.}{ J. S. Bagla, H. K. Jassal and T. Padmamabhan, \textit{Phys. Rev. D}\textbf{67}, 063504(2003).}
\item {12.}{ L. Amendola, M. Quartin, S. Tsujikawa and I. Waga, \textit{Phys.Rev.D}\textbf{74}, 023525(2006).}
\item {13.}{ S. Nojiri, S. D. Odintsov and M. Sasaki, \textit{Phys.Rev. D}\textbf{70} 043539(2004).}
\item {14.}{ C. Wetterich \textit{Nucl. Phys. B}\textbf{302}, 668(1998).}
\item {15.}{ E. J. Copeland, M. Sami and S. Tsujikawa, arXiv:hep-th/060305.}
\item {16.}{ T. Padmanabhan, and T. R. Choudhury, \textit{Phys. Rev. D}\textbf{66}, 081301(2002).}
\item {17.}{ A. Sen, \textit{JHEP} \textbf{0204}, 048(2002).}
\item {18.}{ M. R. Garousi, \textit{Nucl. Phys. B}\textbf{584}, 284(2000).}
\item {19.}{ G. W. Gibbons, \textit{Phys. Lett. B}\textbf{537}, 1(2002).}
\item {20.}{ Y. S. Piao, R. G. Cai, X. m. Zhang and Y. Z. Zhang, \textit{Phys. Rev. D}\textbf{66}, 121301(2002).}
\item {21.}{ L. Kofman and A. Linde, \textit{JHEP} \textbf{0207}, 004(2002).}
\item {22.}{ M. Li, \textit{Phys. Lett. B}\textbf{603} 1(2004), arXiv:hep-th/0403127.}
\item {23.}{ C. Armend\'{a}riz-Pic\'{o}n, V. Mukhanov and P. J. Steinhardt, \textit{Phys.Rev.Lett}\textbf{85}, 4438(2000).}
\item {24.}{ R. J. Sherrer, \textit{Phys. Rev. Lett}\textbf{93)}, 011301(2004).}
\item {25.}{ A. Melchiorri, L. Mersini, C. J. Odman and M. Trodden, \textit{Phys. Rev. D}\textbf{68}, 043509(2003).}
\item {26.}{ T. Chiba, T. Okabe and M. Yamaguchi \textit{Phys. Rev. D}\textbf{62}, 023511(2002)
\\\hspace{0.2 in}S. Capozziello, S. Nojiri and S. D. Odintsov, \textit{Phys. Lett. B}\textbf{632}, 597(2006)}
\item {27.}{ M. Born and Z. Infeld, \textit{Proc. Roy. Soc. A}\textbf{144}, 425(1934).}
\item {28.}{ G. W. Gibbons and C. A. R. Herdeiro, \textit{Phys. Rev. D}\textbf{63}, 064006(2001).}
\item {29.}{ H. Q. Lu, \textit{Int. J. Mod. Phys. D}\textbf{14}, 355(2005);
\\\hspace{0.2 in}W. Fang, H. Q. Lu, Z. G. Huang and K. F. Zhang, \textit{Int. J. Mod. Phys. D}\textbf{15}, 199(2006);
\\\hspace{0.2 in}W. Fang, H. Q. Lu and Z. G. Huang, arXiv:hep-th/0606032.}
\item {30.}{ Z. K. Guo, N. Ohta and Y. Z. Zhang, \textit{Phys. Rev. D}\textbf{72}, 023504(2005);
\\\hspace{0.2 in}Z. K. Guo, N. Ohta and Y. Z. Zhang, \textit{Mod. Phys. Lett. A}\textbf{22}, 883(2007)[arXiv:astro-ph/0603109].}
\item {31.}{ Z. G. Huang, Q. Q. Sun, W. Fang, and H. Q. Lu, \textit{Mod. Phys. Lett. A.}\textbf{22} 3073(2008)[arXiv:hep-th/0612176].}
\item {32.}{ S. Hannestad and E. Mortsell, \textit{Phys. Rev. D}\textbf{66}, 063508(2002).}
\item {33.}{ A. R. Cooray and D. Huterer, \textit{Astrophys. J.}\textbf{513}, L95(1999).}
\item {34.}{ M. Chevallier and D. Polarski, \textit{Int. J. Mod. Phys. D}\textbf{10}, 213(2001). }
\item {35.}{ E. V. Linder, \textit{Phys. Rev. Lett.}\textbf{90}, 091301(2003).}
\item {36.}{ T. Padmanabhan and T.R. Choudhury, \textit{Mon. Not. Roy. Astron. Soc.}\textbf{344}, 823(2003).}
\item {37.}{ B. F. Gerke and G. Efstathiou, \textit{Mon. Not. Roy. Astron. Soc.}\textbf{335}, 33(2002).}
\item {38.}{ S. Hannestad and E. Mortsell, \textit{Phys. Rev. D}\textbf{66}, 063508(2002);
\\\hspace{0.2 in}A. Melchiori, L. Mersini-Houghton, C. J. Odman and M. Trodden, \textit{Phys. Rev. D}\textbf{68}, 043509(2003);
\\\hspace{0.2 in}H. Jassal, J. Bagla and T. Padmanabhan, astro-ph/0506748;
\\\hspace{0.2 in}U. Alam, V. Sahni, T.D. Saini and A.A. Starobinsky, \textit{Mon. Not. R. Astron. Soc.}\textbf{354}, 275(2004)
\\\hspace{0.2 in}S. Capozziello, S. Nojiri, S. D. Odintsov and A. Troisi, \textit{Phys. Lett. B}\textbf{639}, 135(2006).} }
\end{flushleft}
\end{document}